\documentclass[twocolumn,tighten]{aastex62}
\usepackage{amssymb}
\usepackage{array,multirow}
\usepackage{comment}
\usepackage{enumerate}

\newcommand{\gyr}{{\rm{Gyr}}}

\newcommand{\msun}{{M_\odot}}
\newcommand{\rsun}{{R_\odot}}

\newcommand{\metal}{{\rm{Z}}}
\newcommand{\metalsun}{{\rm{Z}_\odot}}
\newcommand{\kpc}{{\rm{kpc}}}

\newcommand{\gaia}{{\it Gaia}}
\newcommand{\cosmic}{{\texttt{COSMIC}}}
 
\newcommand{\days}{\rm{day}}
\newcommand{\yr}{\rm{yr}}

\newcommand{\logg}{$\rm{log\,g}$}
\newcommand{\TESS}{{\it{TESS}}}
\newcommand{\swift}{{\it Swift}}

\newcommand{\mbh}{{M_{\rm{BH}}}}
\newcommand{\porb}{{P_{\rm{orb}}}}
\newcommand{\teff}{{T_{\rm{eff}}}}

\newcommand{\Lx}{{L_{\rm{X}}}}
\newcommand{\Fx}{{F_{\rm{X}}}}
\newcommand{\mas}{\rm{mas}}

\newcommand{\bse}{\texttt{BSE}}

\shorttitle{Constraining Black Hole Formation with 2M05215658+4359220}
\shortauthors{Breivik, Chatterjee, and Andrews}

\begin{document}
\title{Constraining Compact Object Formation with 2M0521}

\author[0000-0002-9660-9085]{Katelyn Breivik}
\affiliation{Canadian Institute for Theoretical Astrophysics, University
of Toronto, 60 St. George Street, Toronto, Ontario, M5S 1A7,
Canada}

\email{kbreivik@cita.utoronto.ca}

\author[0000-0002-3680-2684]{Sourav Chatterjee}
\affiliation{Tata Institute of Fundamental Research, Department of Astronomy and Astrophysics, Homi Bhabha Road, Navy Nagar, Colaba, Mumbai, 400005, India}
\email{souravchatterjee.tifr@gmail.com}

\author[0000-0001-5261-3923]{Jeff J. Andrews}
\affiliation{Foundation for Research and Technology-Hellas, 
100 Nikolaou Plastira St., 
71110 Heraklion, Crete, Greece}
\affiliation{Physics Department \& Institute of Theoretical \& Computational Physics, 
P.O Box 2208, 
71003 Heraklion, Crete, Greece}
\affiliation{DARK, Niels Bohr Institute, University of Copenhagen, 2100 Copenhagen, Denmark}
\email{jeff.andrews@nbi.ku.dk}

\begin{abstract}
We show that the recently discovered binary 2M05215658+4359220 (2M0521), comprised of a giant star (GS) orbiting a suspected black hole (BH) in a $\sim80\,\days$ orbit, may be instrumental in shedding light on uncertain BH-formation physics and can be a test case for studying wind accretion models. Using binary population synthesis with a realistic prescription for the star formation history and metallicity evolution of the Milky Way, we analyze the evolution of binaries containing compact objects (COs) in orbit around GSs with properties similar to 2M0521. We find $\sim10^2-10^3$ CO--GS binaries in the Milky Way observable by \gaia\, and $0-12$ BH--GS and $0-1$ neutron star--GS binaries in the Milky Way with properties similar to 2M0521. We find that all CO--GSs with $\porb<5\,\yr$, including 2M0521, go through a common envelope (CE) and hence form a class of higher mass analogs to white dwarf post-CE binaries. We further show how the component masses of 2M0521-like binaries depend strongly on the supernova-engine model we adopt.
Thus, an improved measurement of the orbit of 2M0521, imminent with \gaia's third data release, will strongly constrain its component masses and as a result inform supernova-engine models widely used in binary population synthesis studies. These results have widespread implications for the origins and properties of CO binaries, especially those detectable by LIGO and LISA. Finally, we show that the reported X-ray non-detection of 2M0521 is a challenge for wind accretion theory, making 2M0521-like CO--GS binaries a prime target for further study with accretion models.  
\end{abstract}


\section{Introduction}\label{S:intro}
Recent discoveries of merging binary black holes (BBHs) and binary neutron stars (BNS) by the LIGO-Virgo observatories \citep[e.g.,][]{Abbott_GW150914,Abbott_GW151226,Abbott_GW170104,Abbott_GW170608,Abbott_GW170814,Abbott_GWcatalog} have reignited widespread interest in the astrophysical origins of compact object (CO) binaries in short-period orbits. One of the major uncertainties in interpreting the observational results  as well as creating predictive models for merger rates and the distributions of expected properties for these sources can be directly attributed to the lack of constraints on quantities related to supernova (SN) physics, more specifically, the CO mass function at birth, and distribution of their natal kicks \citep[e.g.,][]{Chatterjee2017_uncertainties}. Theoretical modeling of the death throes of a massive star is notoriously difficult, and numerical simulations are not yet at a stage to provide strong constraints without observations \citep[e.g.,][]{Fryer2012,Belczynski2012,sukhbold2016,Woosley2017,sukhbold2018,Burrows_etal2018}. 

On the other hand, dark remnants are challenging to discover, hence, it is hard to infer strong constraints from the limited number of discovered COs \citep[e.g.,][]{Gallo2014,Corral-Santana2016}.  
Moreover, the distribution of properties for detected COs suffers from severe selection biases. COs detected in mass-transferring systems detected via their X-ray and radio emissions probe a narrow range in orbital and component properties. 
Similarly, there is strong bias towards detecting distant massive COs via gravitational wave (GW) detectors such as LIGO/Virgo \citep[e.g.,][]{Messenger_Veitch2013}. Given these difficulties, all simulations to estimate merger rates and properties of CO binaries depend critically on prescriptions of the SN-engine models tuned to match the limited number of detected CO binaries \citep[e.g.,][]{Woosley2017,sukhbold2018}. Significant improvement in these prescriptions is possible only via a dramatic increase in the number of detected CO binaries with properties as unbiased as possible. 

While the possibility of identifying COs in detached binaries with luminous companions (LC) was discussed nearly 50 years ago \citep{trimble1969},
astrometric detection of CO--LC binaries has recently come to a sharp focus because of the high expected yield by \gaia\ reported by several independent groups
\citep{Barstow2014,Mashian2017_gaia,Breivik2017_gaia,Yamaguchi2018_gaia,Yalinewich2018_gaia}\footnote{While these studies disagree on the exact number of detached CO--LC binaries \gaia\ will detect in its nominal five-year survey, they all conclude that such binaries exist and \gaia\ should detect many of them.}. Furthermore, since \gaia's detectability of a CO--LC binary depends primarily on its distance and the properties of the companion, the discovered CO population is expected to be less prone to selection biases than the methods discussed earlier. Focusing on BH--LC binaries, \cite{Breivik2017_gaia} showed that the distribution of properties, as well as the expected \gaia\ yield depend on the adopted natal kick distribution. 
Most recently, it has been suggested that detached CO--LCs may also be discovered via photometric variations of the LCs using, for example, \TESS\ data \citep{Masuda2018_TESS}. Since \gaia\ and \TESS\ probe a very different parameter space compared to that covered by X-ray, radio, or GWs, together, they provide 
exciting prospects for both increasing dramatically the known population of COs, as well as exploring a complimentary region of parameter space for CO binaries relative to more traditional detection methods \citep{Breivik2017_gaia}.
CO--LC binaries are being discovered both inside star clusters \citep{2018MNRAS.475L..15G} and in the field \citep[][henceforth T18]{Thompson2018}.
Especially, the recent discovery by T18 that 2M05215658$+$4359220 (henceforth 2M0521) is a giant star (GS) companion to a dark remnant in a $83.2\,\days$ orbit is important in this context since its orbit is expected to be resolved by \gaia.
Combining RV, photometric, and parallactic measurements, T18 proposed that the dark companion to 2M0521 is 
a BH (see \autoref{tab:2M0521} for details).
Our primary goals in this paper are two-fold: First, we study in detail the astrophysical formation channel, Milky Way (MW) abundance, and distribution of properties for detached CO--GS binaries with compact ($P\lesssim5\,\yr$) orbits. 
Second, we investigate the differences in the component masses for the subset of our CO--GS binaries with properties within a narrow range of the observed properties of 2M0521 depending on the adopted SN-engine model.

In \autoref{S:methods} we describe our population synthesis code and adopted SN-engine models. 
In \autoref{S:results} we show our key results which include the dominant formation channel, properties, and abundance for all of MW's CO--GS binaries, and a subset of CO--GS binaries with properties close to those of 2M0521.  
We finish with a summary of our main results in \autoref{S:Conclusion}.

\begin{table}[]
    \centering
    \begin{tabular}{lr}
    \tableline
    \multicolumn{2}{c}{Observed Parameters} \\
    \tableline
    $\porb$ [day] & $83.2\pm0.06$\\
    $ecc$ & $0.0048\pm0.0026$ \\
    $K$ [km s$^{-1}$] & $44.615\pm0.123$ \\
    $\teff$ [K] & $4480\pm62$ \\
    $\logg$ [cgs] & $2.35\pm0.14$ \\
    \tableline
    \multicolumn{2}{c}{Derived Parameters} \\
    \tableline
    Distance [$\kpc$] & 3.11$^{+0.93}_{-0.66}$ \\
    Radius [R$_{\odot}$] & 30$^{+9}_{-6}$ \\
    Luminosity [L$_{\odot}$] & 331$^{+231}_{-127}$ \\
    sin\,$i$ & $0.97^{+0.03}_{-0.12}$ \\
    GS Mass [$\msun$] & $3.2^{+1.0}_{-1.0}$ \\
    CO Mass [$\msun$] & $3.3^{+2.8}_{-0.7}$
    \end{tabular}
    \caption{A summary of the T18 observations and analyses of 2M0521. Fits to radial velocities derived from APOGEE spectra provide the orbital period ($\porb$), eccentricity ($ecc$), and peak radial velocity ($K$), while fits to giant star spectral models provide the temperature ($\teff$). Since APOGEE systematically over estimates surface gravity for stars with large rotational velocity, T18 use TRES spectra to determine log $g$. T18 combine these data with a parallactic distance derived from \textit{Gaia} DR2 and a Monte Carlo analysis of the phased binary motion to fit isochrone models of 2M0521. The best fit values for the radius, luminosity, inclination angle, and component masses are reported.}
    \label{tab:2M0521}
\end{table}

\section{Simulating Milky Way's BH binaries}
\label{S:methods}

We use the population synthesis code \cosmic\footnote{https://cosmic-popsynth.github.io/} to simulate a realistic MW population of CO--GS binaries. 
We adopt a metallicity-dependent MW star formation history, as done in \cite{Lamberts2018}, based on galaxy {\bf{m12i}} in the Latte simulation suite\footnote{The Latte suite of FIRE-2 cosmological, zoom-in, baryonic simulations of MW-mass galaxies \citep{Wetzel2016}, part of the Feedback In Realistic Environments (FIRE) simulation project, were run using the Gizmo gravity plus hydrodynamics code in meshless, finite-mass mode \citep{Hopkins2015} and the FIRE-2 physics model \citep{Hopkins2018}}. 

We assume standard parameter distributions to initialize our binary population. Initial primary masses are distributed according to \citet{Kroupa2001}, and we use a primary-mass-dependent binary fraction \citep[][and references therein]{vanHaaften2013}. Secondary masses are drawn from a flat distribution in mass ratios between $0.001$ and $1$ \citep[e.g.,][]{Mazeh1992}. Orbital periods ($\porb$) are distributed uniformly in log-days \citep{Abt1983}, where the upper bound is $10^5\,\rsun$ and the lower bound is set such that the primary star's radius is less than half of its Roche-lobe radius. We assume a thermal initial eccentricity ($ecc$) distribution \citep{Heggie1975}. 

\cosmic\ uses the binary stellar evolution code \bse\ \citep{Hurley2002} to evolve an initialized population of binaries to the present day. We use standard values and prescriptions adopted by \citet{Hurley2002}, with updates described by \citet{Kiel2008} and \citet{Rodriguez2016}. \bse\ limits binary metallicities to be between $0.005<\metal/\metalsun<1.5$, thus we force all metallicities taken from {\bf{m12i}} to fall within this range. We assume that COs are born with a natal kick drawn from a Maxwellian distribution with $\sigma=265\,\rm{km\,s^{-1}}$ \citep{Hobbs2005}. The natal kicks for BHs are reduced by the fraction of ejected mass that falls back onto the BH during formation \citep{Fryer2012}. We employ the $\alpha\lambda$ CE  prescription where $\alpha=1.0$ is the CE efficiency and $\lambda$ is the non-dimensional binding energy of the stellar envelope, determined with the default \bse\ prescription which has been updated to be dependent on stellar type \citep[][see their appendix]{Claeys2014}. 
We adopt two SN-engine models widely used in binary population synthesis studies \citep{Fryer2012,Belczynski2012}: one model (`rapid') produces a mass gap between BHs and neutron stars (NSs) and the other (`delayed') does not.  The rapid model assumes strong convection which allows instabilities to grow quickly (within $\sim$250\,ms) after core bounce, producing fewer but more energetic SN explosions and BHs with higher masses. The delayed model
allows convective instabilities to grow over a wider range of timescales leading to a continuous distribution of remnant masses.

To scale the number of CO--GSs in our simulations to the MW, we track the total simulated mass, including single and binary stars, required to generate each population of CO--GSs. The total number of CO--GS binaries in a synthetic MW population, $\rm{N}_{\rm{CO-GS}}$, is then the number of simulated CO--GSs normalized by the ratio of the total initial mass of star particles in the {\bf{m12i}} galaxy to the total simulated mass used to generate the population of CO--GS binaries.

From our simulated population, we construct a synthetic present-day Milky Way by sampling the masses, eccentricity, orbital period, metallicity, and age of each CO--GS as well as the temperature and radius of each GS from a multi-dimensional kernel density estimate (KDE). Each sampled binary is assigned an age- and metallicity-dependent position in the galaxy by finding the star particle in {\bf{m12i}} which minimizes the Euclidian distance between the ages and metallicities of the sampled binary and the star particle. Once a star particle is located, we sample the binary's position using an Epanechnikov kernel centered on the selected star particle's location. The kernel widths are taken from \citet{Sanderson2018} who generated synthetic Gaia DR2-like surveys of the Latte suite of FIRE-2 simulations\footnote{Synthetic Gaia DR2-like surveys of the Latte suite of FIRE-2 simulations were created via the Ananke framework \citep{Sanderson2018}}. 
Using the same fixed binary population, we repeat the KDE sampling process 500 times to explore how the population of CO--GS binaries, including those CO--GSs with properties similar to those measured for 2M0521, changes between MW realizations. 

\section{Results}
\label{S:results}

\begin{figure}
    \plotone{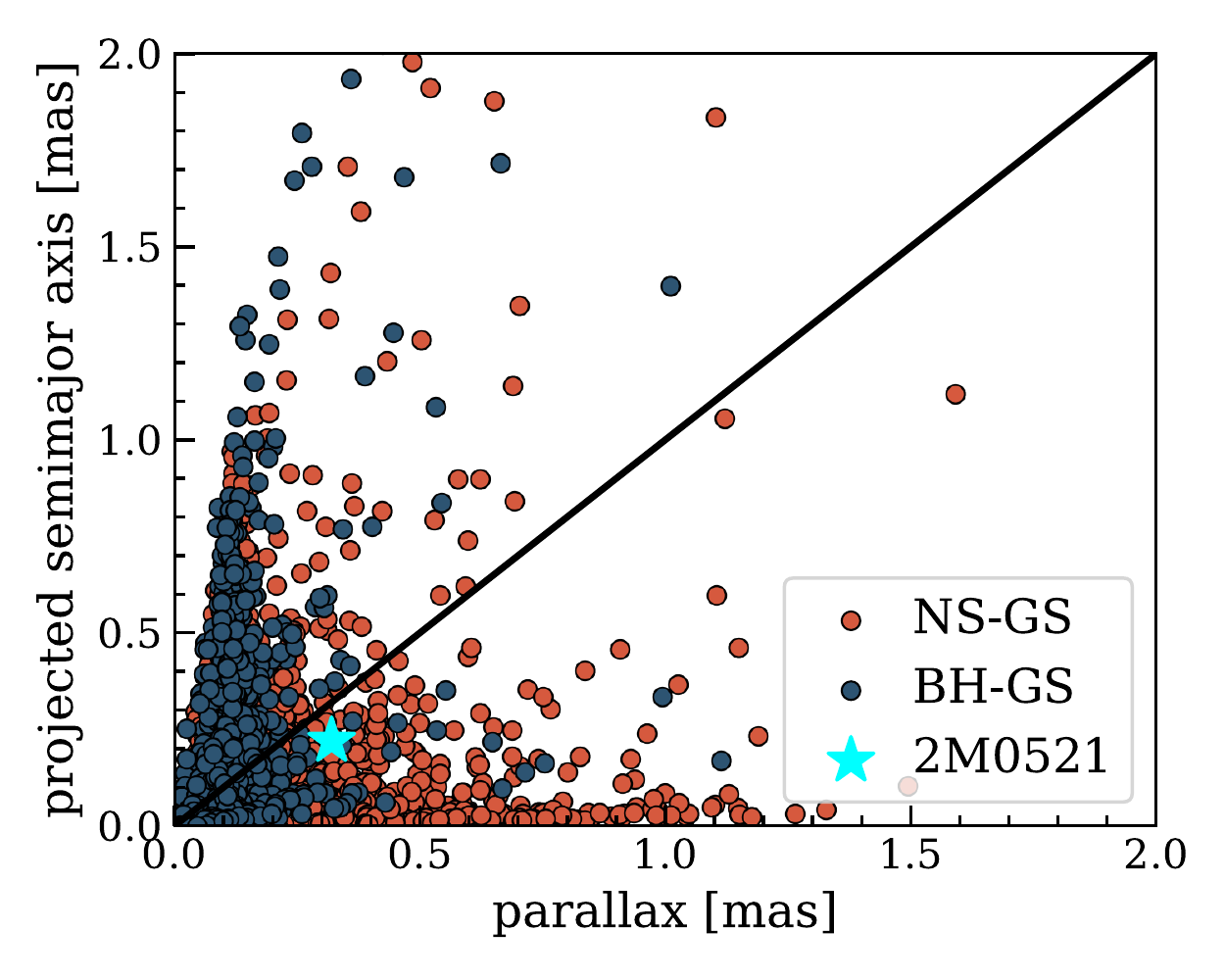}
    \caption{Parallax vs projected semimajor axis for \gaia-detectable NS--GS (orange) and BH--GS (blue) binaries from a single delayed model MW realization. The cyan star shows the results for 2M0521 when using the derived values of T18 from \autoref{tab:2M0521}. Similar results are found for the rapid case. }
    
    \label{fig:parallax}

\end{figure}

\subsection{Contribution of orbital motion to parallax observations}
\label{S:orbit_parallax}
One general result regardless of the CO type or the adopted SN-engine model is that the projected semimajor axis on the sky for CO--GSs detectable by \gaia\ is at least within a factor of two of the parallax for a significant fraction ($\sim15\%$ for NS--GS binaries and $\sim35\%$ for BH--GS binaries) of the CO--GS population.
As an example, Fig.\,\ref{fig:parallax} shows the parallax vs the projected semimajor axis on the sky for a single MW realization using our delayed model of \gaia-detectable CO--GS binaries. Note that for a circular binary the ratio of the projected semimajor axis and the parallax is the true semimajor axis of the binary in au, while for an eccentric binary the ratio is modified by binary orientation including inclination and argument of periastron.

Systems with the largest projected semimajor axes have eccentric orbits with periods of a few years and distances ranging from $3-9\,\kpc$, while systems with the largest parallax always have small ($<1\,\kpc$) distances. 
Based on the properties derived in T18, (\autoref{tab:2M0521}) 
the projected angular size of the semimajor axis at a distance of $3.1\,\kpc$ ($0.22\,\mas$) is a large fraction of the \gaia-measured parallax of $0.27\,\mas$ for 2M0521. This suggests that inferences based on the distance provided by Gaia's parallax without considering binary motion may misclassify CO-GSs, including 2M0521-like binaries, toward lower luminosities and radii. The case of HR 6046, a 5th magnitude star in a six-year orbit with a hidden companion, provides a demonstrative example. The parallax to HR 6046 was measured by {\it{Hipparcos}}; however, \citet{torres2007} showed that this estimate was unreliable and the parallactic motion, proper motion, and astrometric motion all needed to be simultaneously fit to obtain an accurate distance. 

The current \gaia\ parallax uncertainty for 2M0521 is $0.05\,\mas$, which should improve to $0.03\,\mas$\ after five years of observations, suggesting that \gaia\ will be able to astrometrically resolve the orbit of 2M0521 by at least a factor of a few. With the expected inclusion of binary stars in the third data release of \gaia\, an astrometric and radial velocity determination of the orbit for 2M05215 is imminent, thus allowing any degeneracies between the binary motion and parallax motion to be fully understood by \gaia. 

\subsection{Definition of 2M0521-like CO--GS binaries}
We investigate two populations in our results: all detached CO--GSs that are detectable by \gaia\ at present, and the present population of `2M0521-like' CO--GSs. For the \gaia-detectable population, we use the conservative constraints from \citet{Breivik2017_gaia} which require the CO--GS to have an orbital period shorter than the \gaia\ mission length of $5$ years and a projected binary separation that is $3$ times greater than the \gaia\ single pointing position error. 

Likewise, we consider a binary to be 2M0521-like if it fulfills the following additional conditions:
\begin{itemize}
    \item the GS does not fill its Roche radius at present
    \item the orbit is circularized
    \item the GS effective temperature and surface gravity values are within $3\sigma$ of 2M0521's measured values (\autoref{tab:2M0521})
    \item the radial velocity of the GS is at least as large as that of 2M0521, assuming $\sin\,i$=1
\end{itemize}

\begin{figure}
    \centering
    \includegraphics[width=0.47\textwidth]{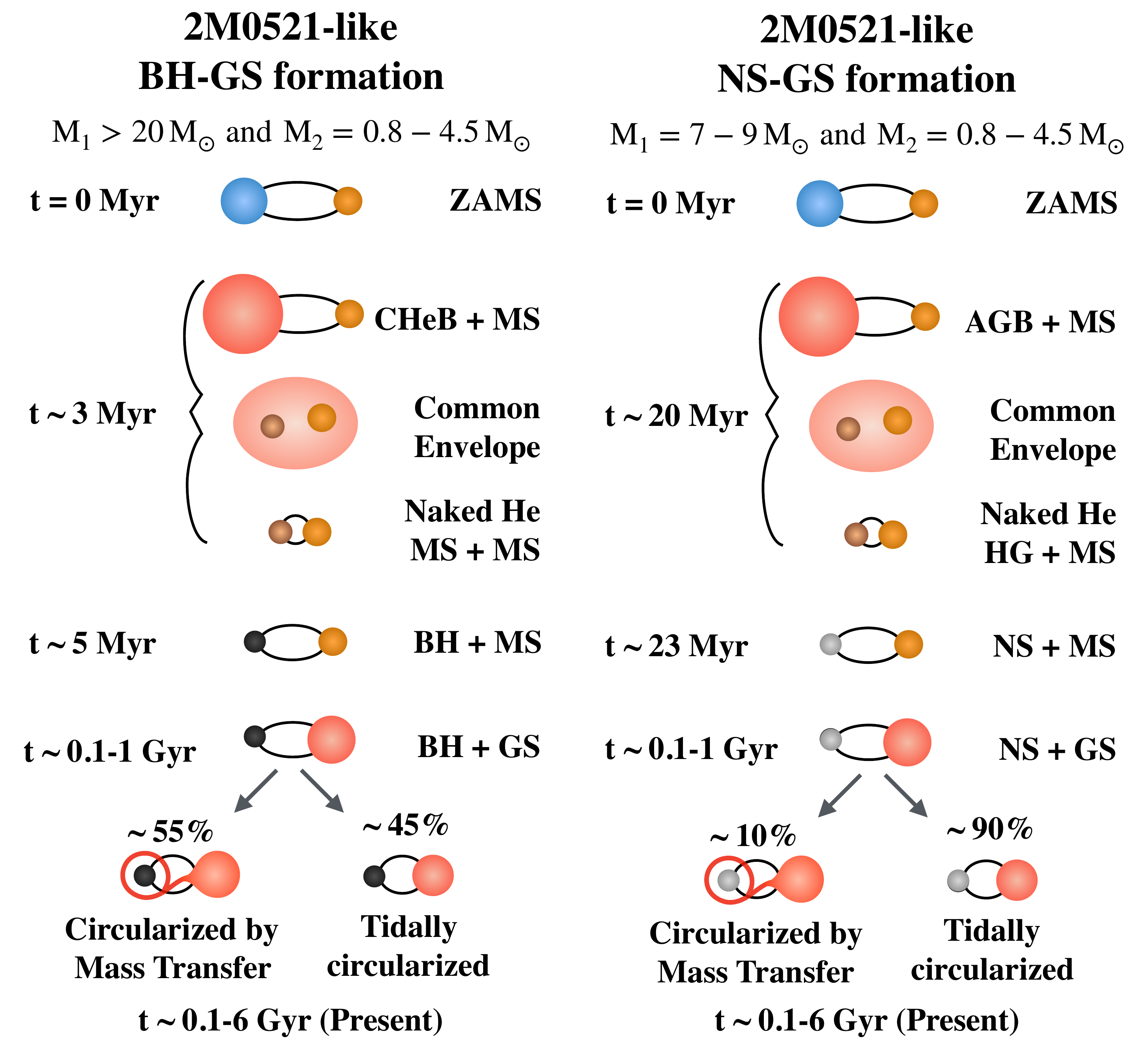}
    \caption{Evolutionary channels which produce 2M0521-like BH--GS or NS--GS binaries. Broadly, the evolution for both channels is similar, requiring a CE phase prior to the compact remnant formation. The differences in the evolution are due to the NS or BH progenitors, which evolve differently based on their initial mass: BH progenitors enter the CE during core helium burning (CHeB) and leave the CE as naked helium MS stars while NS progenitors enter the CE when they are on the asymptotic giant branch (AGB) and leave the CE as naked helium stars on the Hertzsprung Gap.}
    \label{fig:formation_channel}
\end{figure}
\subsection{Formation of compact CO--GS binaries}
\label{subS:formation_channels}

We show examples of the formation scenarios for BH--GS and NS--GS systems that have similar properties to 2M0521 in \autoref{fig:formation_channel}. We find that all CO--GS binaries with $\porb<5\,\yr$, including the 2M0521-like population, must undergo a CE evolution. While the formation scenario stays the same, differences arise between the rapid and delayed models from the different BH masses and relative numbers of BHs and NSs each model produces.

The age of each CO-GS is governed by the evolution time allowed by the GS mass, while the metallicity shows a generally decreasing trend with the time-since-formation of the CO-GS progenitors. The vast majority ($>90\%$) of our present-day CO--GS binaries have super-solar metallicities and ages less than $2\,\gyr$, independent of the SN-engine model. 
These super-solar metallicities lead to most CO progenitors suffering strong line-driven stellar winds \citep{Vink2001} which lead to BH masses that are lower than BHs formed from sub-solar-metallicity progenitors. 

The main difference between the formation channels for BH--GS and NS--GS binaries is set by the evolution timescales of the BH and NS progenitors of the 2M0521-like CO-GSs. The BH progenitors enter a CE with their companion during their core helium burning (CHeB) phase and exit the envelope as naked helium MS stars. In contrast, the lower-mass progenitors of the 
NSs fill their Roche lobes only when they reach the asymptotic giant branch. 
They then leave the CE as naked helium Hertzsprung gap stars. In both cases, the companion star remains a MS through the CE. During the formation of the compact object, a natal kick is applied to the BH or NS which generally increases the eccentricity and semimajor axis of the binary. As the MS companion evolves toward the giant branch, its expanding envelope allows circularization through tides or Roche-lobe overflow mass transfer to take place if the semimajor axis is small enough. A minority of the BH--GS binaries in the delayed model ($<20\%$) and NS--GS binaries in both models ($<5\%$) undergo a second CE when the CO progenitor is on the helium Hertzsrpung Gap. We caution that recent studies \citep[e.g., ][]{Tauris2015} have shown that mass transfer in systems with a helium star donor, the so-called `case BB' mass transfer, remains stable and does not lead to a second CE evolution.

\begin{figure*}
    \plotone{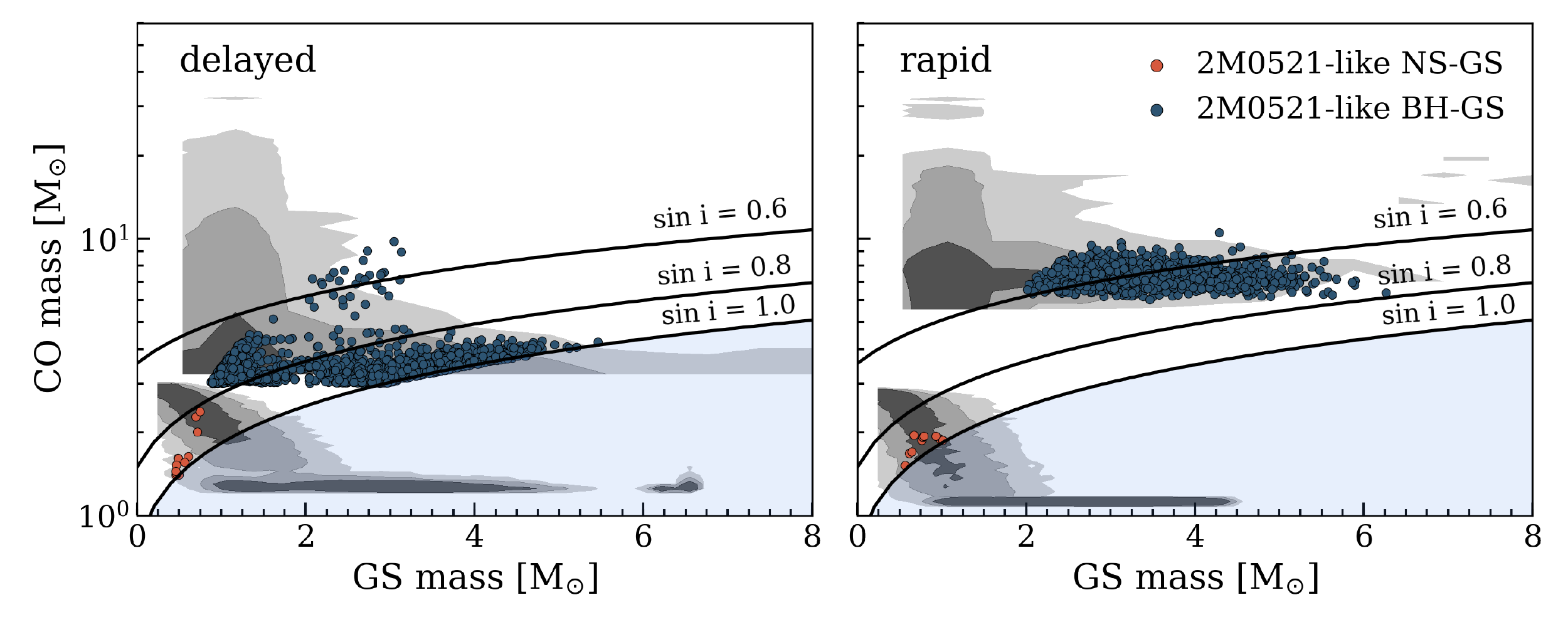}
    \caption{Shaded regions show the $1\sigma$, $2\sigma$, and $3\sigma$ distributions of CO mass vs GS mass for CO--GS binaries for the delayed (left) and rapid (right) SN prescriptions. Orange points denote the 2M0521-like population of NS--GSs and blue points denote the 2M0521-like BH--GSs described in \autoref{subS:formation_channels}. 
    Black lines show contours of constant inclination angle, set by the observed orbital period and radial velocity variations, and the light blue shaded region illustrates the lower bound placed by the radial velocity cuts imposed for the 2M0521-like population with $\sin\,i=1.0$.
    }
    \label{fig:mass_mass}

\end{figure*}

\subsection{Comparing 2M0521-like binaries to \gaia-detectable CO--GS binaries}
\label{subS:BH_formation}
\autoref{fig:mass_mass} shows the distribution of CO and GS masses from 2M0521-like binaries (orange and blue points) and all CO--GS binaries detectable by \gaia\ (gray contours) from our $500$ MW realizations for the rapid and delayed models. The grey contours show the $1\sigma$, $2\sigma$, and $3\sigma$ distributions (descending in shade), where the BH--GS and NS--GS populations are treated independently. Black lines indicate the component mass constraints based on radial velocity observations of 2M0521 for $\sin\,i=0.6,\,0.8\,$ and $1.0$. The imposed constraint $\sin\,i\leq1.0$ rules out systems with masses in the blue shaded region. The overall number of 2M0521-like binaries is heavily weighted toward BH--GS binaries. This is largely the result of the radial velocity constraints we impose based on the observed radial velocity of 2M0521. We note that the small mass gap present in the delayed model panel is due to both the removal of mass transferring NS--GS binaries and the fact that the $3\sigma$ distributions do not contain all CO-GSs. When considering the full population of CO-GS binaries, there is no gap between the lowest mass BHs and the highest mass NSs.

Interestingly, we find that both BHs and NSs are allowed to be the CO in our 2M0521-like populations (\autoref{fig:mass_mass}). Future \gaia\ observations will likely constrain the binary motion, parallax, and inclination of 2M0521 and aid in alleviating the $\sin\,i$ degeneracy. This in turn can put constraints on the nature of the CO and the SN-engine models. For example, an inclination measurement of $\sin\,i<0.8$ rules out NS--GS binaries created using both SN-engine models. Similarly, a measurement of $\sin\,i>0.8$ would make the entire population of BH--GS binaries simulated using the rapid model inconsistent (\autoref{fig:mass_mass}). This is particularly interesting since the best fit model in T18 suggests an inclination of $\sin\,i\sim1$. 
Such constraints are also important since, e.g., the rapid model is often used by several groups in studies which predict the rates and properties of merging binary black hole populations observed by LIGO \citep[e.g., ][]{Belczynski2016, Stevenson2017, Rodriguez2018, Giacobbo2018, Spera2019}. Furthermore, we note that recent population synthesis studies of binary NSs \citep[e.g., ][]{Vigna-Gomez2018} find that the rapid model is inconsistent with current observations with more discoveries of NS binaries needed to distinguish between the delayed model and other SN models.

\begin{figure*}
    \plotone{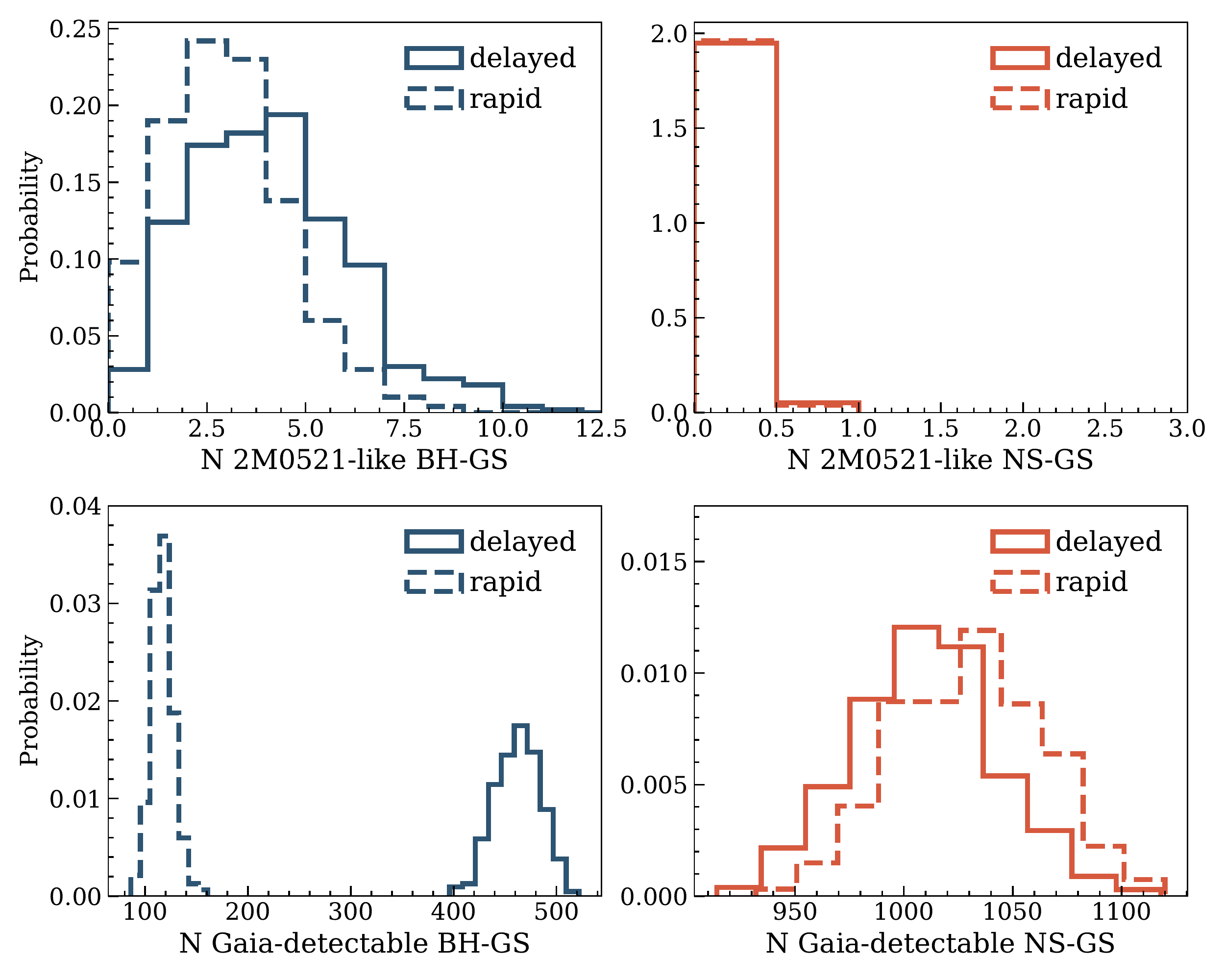}
    \caption{Distribution of the number of 2M0521-like CO--GS binaries (top row) and \gaia-detectable CO--GS binaries (bottom row) from $500$ MW realizations for the delayed and rapid models.} 
    
    \label{fig:numbers}
\end{figure*}

\autoref{fig:numbers} compares the expected numbers per MW for \gaia-detectable CO--GS binaries (bottom) and 2M0521-like binaries (top) formed using rapid and delayed models. 
The increased number of \gaia-detectable NS--GS binaries relative to 2M0521-like NS--GS binaries is a direct consequence of our radial velocity constraints. The higher CO masses of the BH--GS binaries naturally lead to larger radial velocity variations, and are thus less likely to be ruled out. The large numbers of both BH--GS and NS--GS binaries that satisfy our \gaia-detectability cuts suggest that while 2M0521 may be rare, \gaia\ nevertheless has the potential to uncover an interesting class of compact objects that are not widely observed currently. Furthermore, this suggests that astrometric detection of orbital motion by \gaia\ remains one of the most potentially successful methods to observe COs in binaries with complementary properties to those observed through X-ray, radio, and gravitational waves. 
We find a large difference in the expected numbers of \gaia-detectable BH--GS binaries depending on the adopted SN-engine model (\autoref{fig:numbers}). The reason for this is two fold: the rapid model produces roughly two times fewer BHs relative to the delayed model and the relatively lower mass BHs formed in the delayed model receive stronger natal kicks leading to increased BH-GS eccentricities which increase the projected semimajor axes and thus the \gaia\ detectability. This indicates, that \gaia\ is likely to put useful constraints on uncertain SN physics simply from the overall yield of detected BH--GS binaries.

\begin{figure}
   
    \plotone{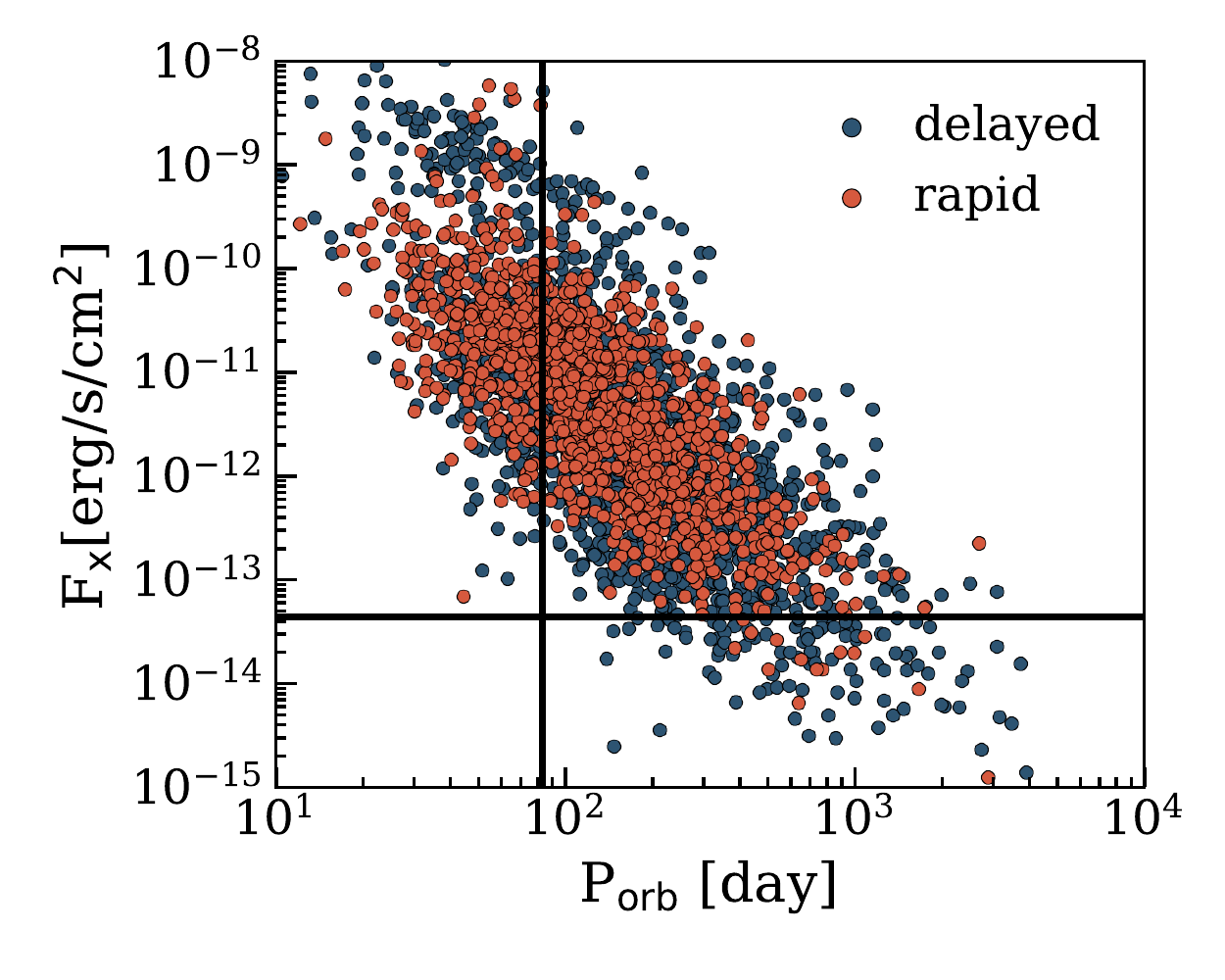}
    \caption{$\Fx$ vs $\porb$ expected from the accretion of the GS wind by the BH for the 2M0521-like BH--GS binaries (see \autoref{SubS:X_rays} for details). The majority of the 2M0521-like BH--GS binaries produce fluxes higher than the upper limit of $4.4\times10^{-14}\,\rm{erg}\, \rm{s}^{-1}\,\rm{cm}^{-2}$ from T18.}
    \label{fig:x_rays}
\end{figure}

\subsection{X-ray Luminosity Puzzle}
\label{SubS:X_rays}

Since the majority of our CO--GS binaries have super-solar metallicities, we expect the GSs to lose mass in winds, which their CO companions can accrete. For NS--GS binaries, the accretion rates are so low that these systems are likely to be within the propeller regime \citep{illarionov1975, ghosh1979}, making them essentially X-ray dark. However, even in the case of low accretion rates, BH--GS binaries can be significant sources of X-ray emission. Following the prescription outlined in \citet{Belczynski2008} for Bondi-Hoyle-Littleton wind accretion, we can derive the X-ray luminosity ($\Lx$) expected from a particular BH--GS binary. 

For instance, using the wind mass loss rate and wind velocity of the well-studied star $\alpha$ Boo (Arcturus; which has properties similar to the GS in 2M0521) artificially placed in an $83.2$-day orbit with a $3\,\msun$ BH, we derive $\Lx\sim10^{35}\rm{erg}\,\rm{s}^{-1}$, nearly four orders of magnitude higher than the X-ray upper limit of 7.7$\times$10$^{31}\rm{erg}\,\rm{s}^{-1}$ calculated from the \swift\ non-detection by T18.

To further investigate this, we applied the accretion formalism of \citet{Belczynski2008} to our simulated binaries, with one adjustment: because the $\dot{M}$ is extremely sub-Eddington, many of these binaries are in the advection-dominated accretion flow (ADAF) regime. We adjust the radiative efficiency of accretion in the ADAF regime following the fitting formulae of \citet{xie12}. Using the simulated positions of these systems in the MW, we convert the calculated X-ray luminosities into X-ray fluxes. \autoref{fig:x_rays} shows that nearly all ($>90\%$) of the 2M0521-like BH--GS binaries are expected to produce X-ray luminosities {\em above} the upper limit calculated by T18 from the \swift\ non-detection.

While these results seem to suggest that the X-ray upper limit is inconsistent with a BH accretor, 
wind accretion in binaries is a notoriously difficult problem \citep{theuns1996, de_val-borro2017} that is complicated by clumping \citep{bozzo2016}, small-scale instabilities \citep{manousakis2015}, and the back reaction of accretion luminosity onto the donor star's atmosphere \citep{sander2018}. Whatever its origin,
the \swift\ X-ray upper limits make 2M0521 an interesting case study for testing wind accretion models and finding the nature of the compact companion. Future, deeper X-ray observations of 2M0521 will place even tighter constraints on a BH accretor. For the entire class of CO--GS binaries detectable by \gaia, eROSITA may be a complementary survey given its X-ray flux limit of $\sim10^{-14}\,\rm{erg}\,\rm{s}^{-1}\rm{cm}^{-2}$ \citep{Clerc2018}.

\section{Conclusion}
\label{S:Conclusion}
We have shown that CO--GS binaries detectable by \gaia\ are naturally produced in binary population synthesis simulations of the MW. One to a few hundreds of BH--GS and roughly a thousand NS--GS binary candidates are expected to be discovered by the end of the five-year \gaia\ mission. While several formation channels produce CO--GS binaries in general, the dominant formation channel for the \gaia-detectable population and those similar to 2M0521 requires evolution through a CE, with a minority of CO--GSs undergoing a second CE. 

We find that the BH masses in the population of 2M0521-like binaries simulated with the rapid model from \citet{Fryer2012} are inconsistent with an inclination with $\sin\,i > 0.8$. Thus, a strong constraint on the inclination, and therefore the mass of the remnant companion to 2M0521 could help to constrain models for the SN-engine and BH formation. Furthermore, if 2M0521's companion is confirmed to be a BH with $\mbh\sim3\,\msun$, the existence of the BH mass gap, and the rapid model can be ruled out altogether. These constraints have wide reaching implications for predictions of BH binaries observed using radio, X-ray, and GWs. 

Finally, we introduced a tension between the non-detection of X-rays from 2M0521 by \swift\ and standard wind accretion theory for the population of 2M0521-like BH--GS binaries. \gaia's third data release is certain to improve the constraints on the mass of 2M0521 and its dark companion and shed light on several of these mysteries.

\acknowledgements
The authors thank the anonymous referee for useful critiques which greatly improved the clarity and quality of this work. K.B. is grateful to Astrid Lamberts for assistance with accessing the Latte simulation suite and to Robyn Sanderson for providing access to the synthetic Gaia DR2 catalogs. J.J.A. acknowledges funding from the European Research Council under the European Union's Seventh Framework Programme (FP/2007-2013)/ERC Grant Agreement n. 617001.

\software{
Astropy \citep{Astropy2013},
COSMIC\,(\doi{10.5281/zenodo.2642803}),
Matplotlib \citep{Hunter2007},
numpy \citep{Van-der-Walt2011},
pandas \citep{McKinney2010},
scipy \url{https://www.scipy.org/}}

\end{document}